\documentclass[prx,superscriptaddress]{revtex4}
\usepackage{color}
\usepackage{amsmath}
\usepackage{amssymb}
\usepackage{graphicx}
\usepackage{tikz}
\usepackage{verbatim}
\usepackage{amsmath}
\usepackage{amssymb}
\usepackage{graphicx}
\usepackage{tikz}
\usepackage{amsmath,bm,amssymb,latexsym,galois,amsthm,euscript}
\usepackage[all,cmtip,knot]{xy}
\usepackage{mathrsfs}
\usepackage[unicode=true,bookmarks=true,bookmarksnumbered=false,bookmarksopen=false,breaklinks=false,pdfborder={0 0 1},backref=false,colorlinks=true]{hyperref}
\usepackage{xcolor}
\usepackage{overpic}
\usepackage{algpseudocode}
\usepackage{algorithm}

\hypersetup{
    colorlinks,
    linkcolor={red!50!black},
    citecolor={blue!90!black},
    urlcolor={blue!90!black}
}

\makeatletter
\@ifundefined{textcolor}{}
{%
 \definecolor{BLACK}{gray}{0}
 \definecolor{WHITE}{gray}{1}
 \definecolor{RED}{rgb}{1,0,0}
 \definecolor{GREEN}{rgb}{0,1,0}
 \definecolor{BLUE}{rgb}{0,0,1}
 \definecolor{CYAN}{cmyk}{1,0,0,0}
 \definecolor{MAGENTA}{cmyk}{0,1,0,0}
 \definecolor{YELLOW}{cmyk}{0,0,1,0}
}

\usepackage{amsfonts}\usepackage{tabularx}\usepackage{dcolumn}\usepackage{bm}\usepackage{graphicx}\usepackage{epstopdf}

\hypersetup{urlcolor=blue}

\newcommand{\bv}{\mathbf{v}}
\newcommand{\ketv}{|\mathbf{v}\rangle}
\newcommand{\ketPsi}{|\Psi\rangle}
\newcommand{\tT}{\tilde{T}}

\usepackage{amsthm}

\theoremstyle{definition}
\newtheorem{defn}{Definition}
\newtheorem{Problem}[defn]{Problem}
\newtheorem{Example}[defn]{Example}
\theoremstyle{remark}

\makeatother

\begin{document}

\title{An Efficient Algorithmic Way to Construct Boltzmann Machine Representations for Arbitrary Stabilizer Code}

\author{Yuan-Hang Zhang}
\affiliation{Department of Physics, University of California, San Diego, CA 92093, USA}
\affiliation{School of the Gifted Young, University of Science and Technology of China, Hefei, Anhui, 230026, P.R. China}

\author{Zhian Jia}
\email{giannjia@foxmail.com}
\affiliation{Centre for Quantum Technologies, National University of Singapore, Singapore 117543, Singapore}
\affiliation{CAS Key Laboratory of Quantum Information, School of Physics, University of Science and Technology of China, Hefei, Anhui, 230026, P.R. China}
\affiliation{CAS Center For Excellence in Quantum Information and Quantum Physics, University of Science and Technology of China, Hefei, Anhui, 230026, P.R. China}
\affiliation{Microsoft Station Q and Department of Mathematics, University of California, Santa Barbara, California 93106-6105, USA}

\author{Yu-Chun Wu}
\email{wuyuchun@ustc.edu.cn}
\affiliation{CAS Key Laboratory of Quantum Information, School of Physics, University of Science and Technology of China, Hefei, Anhui, 230026, P.R. China}
\affiliation{CAS Center For Excellence in Quantum Information and Quantum Physics, University of Science and Technology of China, Hefei, Anhui, 230026, P.R. China}

\author{Guang-Can Guo}
\affiliation{CAS Key Laboratory of Quantum Information, School of Physics, University of Science and Technology of China, Hefei, Anhui, 230026, P.R. China}
\affiliation{CAS Center For Excellence in Quantum Information and Quantum Physics, University of Science and Technology of China, Hefei, Anhui, 230026, P.R. China}

\begin{abstract}
Restricted Boltzmann machines (RBMs) have demonstrated considerable success as variational quantum states; however, their representational power remains incompletely understood. In this work, we present an analytical proof that RBMs can exactly and efficiently represent stabilizer code states—a class of highly entangled quantum states that are central to quantum error correction. Given a set of stabilizer generators, we develop an efficient algorithm to determine both the RBM architecture and the exact values of its parameters. Our findings provide new insights into the expressive power of RBMs, highlighting their capability to encode highly entangled states, and may serve as a useful tool for the classical simulation of quantum error-correcting codes.\end{abstract}
\maketitle

\section{Introduction}
To conquer one of the main challenges, the dimensionality problem (also known as Hamiltonian complexity \cite{osborne2012hamiltonian,verstraete2015quantum}),
 in condensed matter physics, many different representations of quantum many-body states are developed. 
 For example, the well-known tensor network representations \cite{Friesdorf2015,Orus2014tensornet,Verstraete2009} including density-matrix renormalization group (DMRG) \cite{White1992}, matrix product states (MPS) \cite{verstraete2008matrix}, projected entangled pair states (PEPS) \cite{verstraete2004renormalization,verstraete2008matrix}, folding algorithm \cite{Banuls2009}, entanglement renormalization \cite{Vidal2007}, time-evolving block decimation (TEBD) \cite{Vidal2003} and string-bond state \cite{Schuch2008} \emph{etc}., have gradually became a standard method in solving quantum many-body problems. The efficiency of the tensor network representations is known to be partially based on the entanglement properties of the state.

In recent years, a new representation based on a shallow neural network—the restricted Boltzmann machine (RBM)—was introduced by Carleo and Troyer \cite{Carleo602}. They demonstrated the expressive power of this representation by computing the ground states and unitary dynamics of the transverse-field Ising model and the antiferromagnetic Heisenberg model. 
Subsequently, various aspects of the RBM representation have been explored. 
Deng \emph{et al.} analyzed the entanglement properties of RBM states \cite{Deng2017}, while Gao and Duan extended the framework to deep Boltzmann machines (DBM) \cite{gao2017efficient}, and the entanglement property of DBM state is given in \cite{jia2020entanglement}. 
The connections between tensor networks and RBM representations have also been investigated in Refs.~\cite{gao2017efficient,huang2017neural,Chen2018,Glasser2018}. 
Moreover, many other neural network architectures, like convolutional neural network, transformer neural network, etc., have also been proposed for the efficient representation of quantum many-body states (and density operator), see e.g., \cite{Liang2018,Zhang2023Transformer,wei2025variational,Torlai2018latent, Hartmann2019neural, Nagy2019variational}; see Refs.~\cite{Carleo2019machine,jia2019quantum,lange2024architectures,melko2019restricted} for comprehensive reviews and outlooks.

A central problem in the study of RBM states is to understand their representational power. Although the universality of RBMs has long been established \cite{le2008representational}, the number of hidden neurons required to represent an arbitrary distribution generally scales exponentially, rendering such constructions impractical. 
For RBM-based quantum states, despite numerous numerical investigations, analytical results remain scarce.

Notable analytical results on RBM representations for specific quantum states include the toric code state \cite{Deng2017a}, the one-dimensional symmetry-protected topological (SPT) cluster state \cite{Deng2017a}, and the graph state \cite{gao2017efficient}. In Ref.~\cite{jia2018efficient}, we investigated the RBM representation of the planar code within the stabilizer formalism and explicitly constructed sparse RBM architectures for specific stabilizer groups. However, a general construction remains elusive. Since the publication of the preprint version of this work, several studies have further explored neural network representations of code states of stabilizer codes (or equivalently ground state of local commutative Hamiltonian); see, for example, Refs.~\cite{pei2021compact,Chen2025RBM,zheng2019RBM}.

In this work, we comprehensively investigate the RBM representation for stabilizer code states \cite{gottesman1997stabilizer,Gottesman1998} which cover a large classes of states studied before. 
An algorithmic way to construct RBM parameters of an arbitrary stabilizer group is given, which gives a complete solution of the problem of understanding the representational power of RBM in the stabilizer formalism.

The paper is organized as follows. In Section~\ref{sec:pre} and Section~\ref{sec:code}, we begin by reviewing some basic concepts of RBM states and stabilizer codes. In Section~\ref{sec:RBMcode}, we provide a detailed discussion on how to construct an RBM state for a given stabilizer code. The final section offers concluding remarks and an outlook for future work.

\section{Preliminary Notions}
\label{sec:pre}

\begin{figure}[t]
    \centering
    \includegraphics[width=0.6\linewidth]{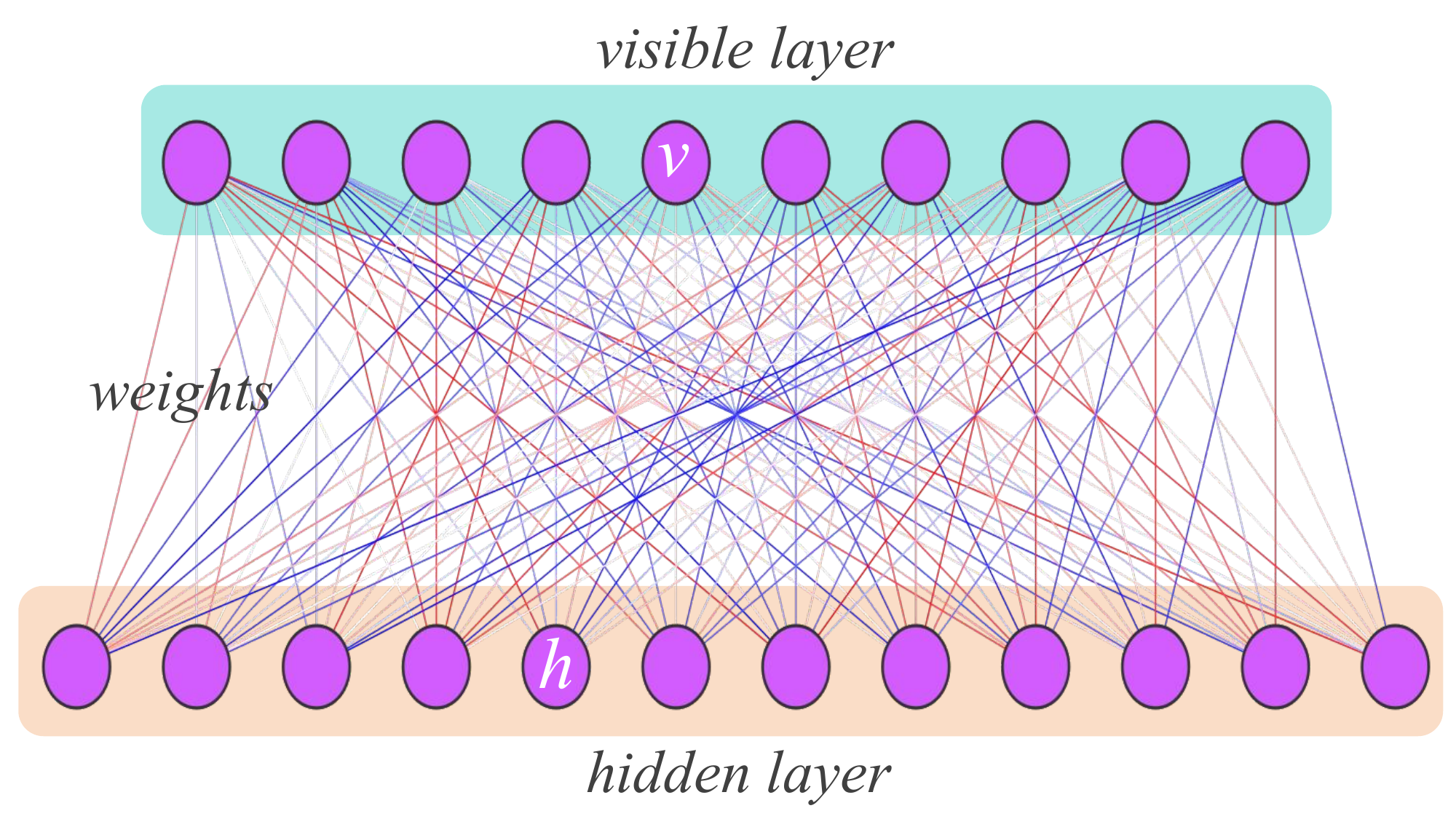}
    \caption{Illustration of an RBM state. The RBM network consists of a visible layer and a hidden layer. Neurons within the same layer are not connected, while connections exist between neurons in different layers. The visible neurons $\mathbf{v}=(v_1,\cdots,v_n)$ represent the basis states of a given Hilbert space, the output of the RBM represents the coefficients $\Psi_{RBM}(\mathbf{v})$ for each basis.
    \label{fig:RBMstate}}
\end{figure}

An RBM is a two-layer neural network consisting of $n$ visible neurons $v_i \in \{0, 1\}^n$ and $m$ hidden neurons $h_j \in \{0, 1\}^m$; see Figure~\ref{fig:RBMstate}. The connections between the visible and hidden layers are specified by a weight matrix $W_{ij}$, while $a_i$ and $b_j$ denote the bias terms for the visible and hidden neurons, respectively. These parameters define a joint probability distribution over visible and hidden units:
\begin{equation}
    p(\mathbf{v}, \mathbf{h}) = \frac{1}{Z} \exp\left( \sum_i a_i v_i + \sum_j b_j h_j + \sum_{i,j} W_{ij} v_i h_j \right),
\end{equation}
where the partition function $Z$ is given by
\begin{equation}
    Z = \sum_{\mathbf{v}, \mathbf{h}} \exp\left( \sum_i a_i v_i + \sum_j b_j h_j + \sum_{i,j} W_{ij} v_i h_j \right).
\end{equation}

To represent a quantum many-body state, we map the local degrees of freedom of the quantum system onto the visible neurons and trace out the hidden neurons, yielding
\begin{equation}
\begin{aligned}
    \psi_{\mathrm{RBM}}(\mathbf{v}) &= \sum_{\mathbf{h}} p(\mathbf{v}, \mathbf{h}) \\
    &= \frac{1}{Z} \exp\left( \sum_i a_i v_i \right) \prod_{j} \left( 1 + \exp\left( b_j + \sum_i v_i W_{ij} \right) \right), \\
    |\Psi_{\mathrm{RBM}}\rangle &= \sum_{\mathbf{v}} \psi_{\mathrm{RBM}}(\mathbf{v}) |\mathbf{v}\rangle.
\end{aligned}
\end{equation}
The parameters of the RBM can be chosen as complex numbers, in which case the resulting joint distribution $p(\mathbf{v}, \mathbf{h})$ may take complex values. For more details, see, e.g., Refs.~\cite{Carleo602, jia2018efficient, jia2019quantum}.

A stabilizer group $\mathbf{S}$ is defined as an Abelian subgroup of the Pauli group $\mathbf{P}_n=\{I,X,Y,Z\}^{\otimes n}\times\{\pm 1, \pm i\}$ that stabilizes an invariant subspace $\mathcal{C}$ of the total space $\mathcal{H}=(\mathbb{C}^2)^{\otimes n}$ with $n$ physical qubits. The space $\mathcal{C}$ is called the code space of the stabilizer group $\mathbf{S}$. More precisely, $\forall T \in \mathbf{S}, \forall |\Psi\rangle\in\mathcal{C}$, the equation $T|\Psi\rangle=|\Psi\rangle$ is always satisfied. Suppose $\mathbf{S}$ is generated by $m$ independent operators,  $\mathbf{S}=\langle T_1,T_2,\cdots,T_m\rangle$. It is easy to check the following properties for the stabilizer operators:
\begin{enumerate}
\item $T_j^2=I$ for all $j$, $[T_i,T_j]=0$ and $-I,\pm i I\not\in\mathbf{S}$.
\item $\langle T_1,\cdots,T_k,\cdots, T_m\rangle=\langle T_1,\cdots,T_j T_k,\cdots, T_m\rangle$, for any $j\neq k$.
\end{enumerate}
We refer the reader to Ref.~\cite{Nielsen2010} for more details about stabilizer code.

Our goal is to find the RBM representation of code states $|\Psi_L\rangle \in \mathcal{C}$. We present an explicit algorithm to construct a set of basis code states $\{|\Psi_L\rangle\}$ that span the code space $\mathcal{C}$ for an arbitrary stabilizer group. To summarize, we aim to address the following problem:

\begin{Problem}
Given a stabilizer group $\mathbf{S}$ generated by $m$ independent stabilizer operators $T_1, \dots, T_m$, does there exist an efficient RBM representation of the code states $|\Psi_L\rangle$? If so, how can one determine the corresponding RBM parameters?
\end{Problem}

To answer this, we first introduce the standard form of a stabilizer code~\cite{gottesman1997stabilizer, Gottesman1998, Nielsen2010}.

\section{Standard Form of Stabilizer Code}
\label{sec:code}

Every Pauli operator $T_k$ that squares to identity can be written as  $\alpha_k P(x_1,z_1)\otimes P(x_2,z_2)\otimes\cdots\otimes P(x_n,z_n)$, where  $\alpha_k=\pm1$ is a phase factor, and $P(x_i,z_i)$ is one of the Pauli matrices,
\begin{equation*}
P(x_i,z_i)=\left\{
\begin{array}{ll}
I\ &\mathrm{if}\ x_i=0, z_i=0\\
X\ &\mathrm{if}\ x_i=1, z_i=0\\
Y\ &\mathrm{if}\ x_i=1, z_i=1\\
Z\ &\mathrm{if}\ x_i=0, z_i=1
\end{array}\right.
\end{equation*}
In this way, every stabilizer operator $T_k$ can be written as the combination of a phase factor $\alpha_k$ and a binary vector $a_k=(x_1, \cdots, x_n, z_1, \cdots, z_n)$. It is easy to prove that if $T_l=T_j T_k$, then $a_l=a_j\oplus a_k$, where $\oplus$ denotes for bitwise addition modulo 2.

For the set of stabilizer generators $\{T_1,\cdots, T_m\}$, we can stack all the binary vectors $a_k$ together to form an $m\times 2n$ matrix $A$, called the check matrix \cite{Nielsen2010}. Each row of $A$ is a vector $a_k$ that corresponds to a stabilizer operator $T_k$. {To clarify the notation, we denote $A=(A_X|A_Z)$, where $A_X$ and $A_Z$ are $m\times n$ matrices denoting the $x$ and $z$ part of the binary vector $a$, respectively.} Since $\{T_1,\cdots, T_k, \cdots, T_m\}$ and $\{T_1,\cdots, T_jT_k, \cdots, T_m\}$ generates the same stabilizer group, one can add one row of $A$ to another row of $A$ (modulo 2) without changing the code space $\mathcal{C}$. Meanwhile, {swapping the $p$-th and $q$-th row of $A$ corresponds to relabeling the stabilizer generators $T_p\leftrightarrow T_q$, and simultaneously swapping the $p$-th and $q$-th column of both $A_X$ and $A_Z$ corresponds to relabeling the qubits $v_p\leftrightarrow v_q$.} 

{With the adding and swapping operations,} we can perform \emph{Gaussian elimination} to the check matrix $A$ \cite{Nielsen2010, yan2012k}. This leads us to the standard form of a stabilizer group. This procedure is well-established, and one can refer to \cite{Nielsen2010} for a detailed explanation. Here, we briefly outline the routine and clarify the notation:  

We start from the original check matrix $A=(A_X|A_Z)$. Performing Gaussian elimination to $A_X$, we obtain

\begin{equation}
\left(\begin{array}{cc|cc}
I_p & B & C & D\\
0   & 0 & E & F
\end{array}\right)
\end{equation}
where $I_p$ is a $p\times p$ identity. Note that we must keep track of the phase factors $\{\alpha_1, \cdots, \alpha_m\}$ during this procedure. Further performing Gaussian elimination to $F$, we can get another identity $I_q$:

\begin{equation}
\left(\begin{array}{ccc|ccc}
I_p & B_1 & B_2 & C   & D_1 & D_2\\
0   & 0   & 0   & E_1 & I_q & F_1
\end{array}\right)
\end{equation}
Finally, we can use $I_q$ to eliminate $D_1$:

\begin{equation}\label{eq:std}
\left(\begin{array}{ccc|ccc}
I_p & B_1 & B_2 & C_1 & 0   & D_3\\
0   & 0   & 0   & E_1 & I_q & F_1
\end{array}\right)
\end{equation}

Eq. (\ref{eq:std}) is called the standard form of a stabilizer code. There are $p+q$ independent stabilizer generators, and the number of qubits encoded is $k=n-p-q$. If the stabilizer generators we start with are not independent with each other, zero rows will be encountered during the elimination, which we can simply discard and finally reaching a set of independent generators.

One advantage of Eq. (\ref{eq:std}) is that it is easy to construct logical $X$ and $Z$ operators from it. The check matrix for logical $X$ and $Z$ operators, $\bar{X}$ and $\bar{Z}$, can be chosen as:
\begin{equation}
\begin{aligned}
    &A_{\bar{X}}=(0\ F_1^T\ I|D_3^T\ 0\ 0),\\
    &A_{\bar{Z}}=(0\ 0\ 0|B_2^T\ 0\ I).
\end{aligned}\label{eq:logicalop}
\end{equation}
One can verify that these operators all commute with the stabilizer generators and commute with each other except that $\bar{X}_j$ anti-commutes with $\bar{Z}_j$ \cite{Nielsen2010}.

\begin{algorithm}[ht]
\caption{Constructing RBM representation for the code state of the stabilizer group $\mathbf{S}=\langle T_1,\cdots,T_n\rangle$}
\label{alg:RBMstabilizer}
\begin{algorithmic}[1]
\Require  A set of $n$ commuting stabilizer generators $G=\{T_1,\dots ,T_n\}$ acting on $n$ qubits.
\Ensure   RBM parameters $\Omega=\{a_i,\;b_\mu,\;W_{i\mu}\}$ that realize the corresponding code state.

\Statex{\bf Initialize}
    \State Set all visible biases $a_i\gets 0$.
    \State Set the (temporary) visible–visible matrix $J_{ij}\gets 0$.
    \State Set hidden–layer size $n_h\gets 0$ and initialise $b_\mu,W_{i\mu}$ to $0$.

\Statex{\bf Bring the generators into standard form}
    \State Use Gaussian elimination to put $G$ into the check-matrix form
          $\bigl(I_p\;B\;|\;C\;0\bigr)\oplus\bigl(0\;0\;|\;E\;I_r\bigr)$, keeping track of the phases $\{\alpha_1, \cdots, \alpha_n\}$.

\Statex{\bf X–type generators $\bigl\{\tilde T_1,\dots ,\tilde T_p\bigr\}$}
    \For{$j=1$ \textbf{to} $p$}
    \State $a_j\gets a_j+\log \alpha_j$
        \Comment{$\tilde T_j$ flips qubit $j$ only}
        \For{$k=1$ \textbf{to} $j-1$}
            \If{$C_{jk}=1$}           \Comment{$Z$ on earlier qubit $k$}
                \State $J_{jk}\gets J_{jk}+i\pi$
            \EndIf
        \EndFor
    \EndFor

\Statex{\bf Z–type generators $\bigl\{\tilde T_{p+1},\dots ,\tilde T_{n}\bigr\}$}
    \For{$\ell=p+1$ \textbf{to} $n$}
        \State Create a hidden neuron $h_\ell$; set $n_h\gets n_h+1$
        \State $b_\ell\gets \log \alpha_\ell, W_{\ell\ell}\gets i\pi$
        \ForAll{$k$ with $E_{\ell k}=1$}  \Comment{$Z$ on qubit $k$}
            \State $W_{k\ell}\gets i\pi$
        \EndFor
    \EndFor

\Statex{\bf Remove visible–visible couplings}
    \ForAll{$(j,k)$ with $J_{jk}\neq 0$ and $j<k$}
        \State Introduce a new hidden neuron $h$ 
        \State Compute $a_j$, $a_k$, $b_h$, $W_{jh}$, $W_{kh}$ according to Eqs.~(\ref{eq:vconn})–(\ref{eq:sol})
        \State $J_{jk}\gets 0$
    \EndFor

\State \Return $\Omega=\bigl\{a_i,\;b_\mu,\;W_{i\mu}\bigr\}$
\end{algorithmic}
\end{algorithm}

\section{RBM Representation for an Arbitrary Stabilizer Group}
\label{sec:RBMcode}

In this section, we illustrate how to construct the RBM representation for any given stabilizer group.

Suppose the set of stabilizer generators have already been brought into the standard form like Eq. (\ref{eq:std}). To begin with, we need to specify one code state in the code space $\mathcal{C}$. As an example, we choose the logical $Z$ eigenstate with eigenvalue $1$, i.e., $\bar{Z}_i|\Psi\rangle=|\Psi\rangle, 1\leq i\leq k$. We can see that we are actually treating the logical $Z$ operators $\bar{Z}_i$ as new independent stabilizer operators, and the stabilized subspace is narrowed down to containing one state only. The set of independent stabilizer generators now becomes $\{T_1, \cdots, T_m, \bar{Z}_1, \cdots, \bar{Z}_k\}$, with the new check matrix being

\begin{equation}\label{eq:stack}
\left(\begin{array}{ccc|ccc}
I_p & B_1 & B_2 & C_1 & 0   & D_3\\
0   & 0   & 0   & E_1 & I_q & F_1\\
0   & 0   & 0   & G   & 0   & I_k
\end{array}\right)
\end{equation}

Upon introducing new independent stabilizer operators, Eq. (\ref{eq:stack}) can be further simplified. Eliminating $D_3$ and $F_1$ with $I_k$, we obtain the final form of the check matrix:

\begin{equation}\label{eq:final}
\left(\begin{array}{cc|cc}
I_p & B & C & 0\\
0   & 0 & E & I_r
\end{array}\right)
\end{equation}
where $r=q+k$, and $p+r=n$. Denote the $n$ stabilizer generators corresponding to Eq. (\ref{eq:final}) as $\{\tilde{T}_1,\cdots, \tilde{T}_n\}$. We call $\tilde{T}_1,\cdots,\tilde{T}_p$ X-type stabilizers, denoted by $\tilde{T}^x$, and $\tilde{T}_{p+1},\cdots,\tilde{T}_n$ Z-type stabilizers, denoted by $\tilde{T}^z$.

With $n$ qubits, the $n$ stabilizer operators uniquely determines one stabilizer code state. Next, we explicitly construct this state and then translate it into the parameters of an RBM.

We start with the full expansion of the quantum state in the Pauli-Z basis
\begin{equation}
    \ketPsi=\sum_{\bv\in\{0, 1\}^n}\psi(\bv)\ketv
\end{equation}
where $\bv=(v_1, v_2, \cdots, v_n)$. 

Because $\ketPsi$ is a stabilizer state, every generator $\tT_i$ satisfies $\tT_i \ketPsi=\ketPsi$. Applying $\tT_i$ on the expansion gives
\begin{equation}
    \sum_{\bv}\psi(\bv)\ketv = \sum_{\bv}\psi(\bv)\tT_i\ketv \label{eq:expansion}
\end{equation}

To begin with, we consider a Z-type stabilizer $\tT^z_i, i\in\{p+1, \cdots, n\}$. By construction, it contains only the single-qubit operators $Z$ or $I$ and an overall phase $\alpha_i$. Therefore, it is diagonal in the computational basis: 

\begin{equation}
    \tT^z_i\ketv=\phi_i(\bv)\ketv,\quad \phi_i(\bv)=\alpha_i(-1)^{\sum_{j:\ Z_j\in \tT_z}v_j}\in\{+1, -1\}
\end{equation}

Plugging this into Eq.~\eqref{eq:expansion} we get
\begin{equation}
    \psi(\bv)=\phi_i(\bv)\psi(\bv)
\end{equation}

Hence, for every basis string either $\phi_i(\bv)=+1$, in which case $\psi(\bv)$ may be non-zero; or $\phi_i(\bv)=-1$, in which case $\psi(\bv)$ must vanish. 
Each Z-type stabilizer therefore imposes a simple parity rule on the bit string: only those $\bv$ whose selected qubits sum to an even (or odd) parity survive.

After Gaussian elimination, the full check matrix has the block form, Eq.~\eqref{eq:final}, and the bottom-right block corresponds to the $r=n-p$ Z-type stabilizers. 

Crucially, each of these $r$ generators acts with a single $Z$ on a distinct one of the last $r$ qubits. Because of this, once the first $p$ qubits $(v_1, \cdots, v_p)$ are fixed, the parity constraints from the Z-type stabilizers uniquely determine the remaining $r$ bits. 

Using the notation of the check matrix
\begin{equation}
    \big(0\ 0\ |\ E\ I_r\big),
\end{equation}
the parity constraint can be expressed as
\begin{equation}
\begin{aligned}
    d_j =& \Big(1 + \alpha_j (-1)^{\sum_k E_{jk}v_k}\Big) / 2\\
    =& \Big(1 + \exp(\log \alpha_j + i\pi\sum_k E_{jk}v_k)\Big) / 2\\
    =&
    \begin{cases}
    1, & \text{constraint satisfied,}\\
    0, & \text{constraint violated.}
    \end{cases}
    \label{eq:Z_full}    
\end{aligned}
\end{equation}

In an RBM, Eq.~\eqref{eq:Z_full} can be implemented by adding a hidden neuron $h_j$ with bias $\log \alpha_j$ and weight $W_{kj}=i\pi$ between $v_k$ and $h_j$. 

We will therefore label any basis state that does satisfy all Z-type constraints as 
\begin{equation}
    |v_1v_2\cdots v_p \times\cdots\times\rangle,
\end{equation}
where the ``$\times$'' symbols stand for the uniquely determined values of qubits $p+1, \cdots, n$. All other computational-basis states have amplitude zero. 

We have already seen that the $r=n-p$ Z-type generators keep only those basis strings $|v_1v_2\cdots v_p \times\cdots\times\rangle$ that satisfy their parity rules. What remains is to determine the complex amplitude attached to each of those surviving strings. That information is contained in the $p$ X-type stabilizers $\tT^x_1, \cdots, \tT^x_p$. 

In the reduced check matrix
\begin{equation}
    \big(I_p\ B\ |\ C\ 0\big),
\end{equation}
\begin{itemize}
    \item The identity block $I_p$ says that $\tT^x_j$ flips only qubit $j$;
    \item The matrix $C$ tells us which qubits carry an extra $Z$ in $\tT^x_j$;
    \item The matrix $B$ flips the ``$\times$'' qubits that are uniquely determined by the Z-type stabilizers. As stabilizers commute, the flips are consistent with the parity constraints, and we can ignore them for now;
    \item The zeros in the last $r$ columns guarantee that none of the ``$\times$'' qubits contributes any phase.
\end{itemize}

Because all generators commute, we can build any allowed basis string by starting from the ``all-zero'' string
\begin{equation}
    |00\cdots 0 \times\cdots\times\rangle,
\end{equation}
and applying $\tT^x_j$ exactly when $v_j=1$. 

Acting with $\tT^x_j$ flips the bit $v_j\to\bar{v}_j=1-v_j$ and multiplies the state by a phase
\begin{equation}
    c_j(\bv)=\alpha_j (-1)^{\sum_{k}C_{jk}v_k} = \exp\big(\log\alpha_j + i\pi\sum_{k}C_{jk}v_k\big)\in\{\pm1, \pm i\}
\end{equation}
where $\alpha_j\in\{\pm1, \pm i\}$ is the overall phase appearing in $\tT^x_j$. 

Because $\tT^x_j\ketPsi=\ketPsi$, we must have
\begin{equation}
    \psi(v_1\cdots \bar{v}_j \cdots v_p \times\cdots\times) = c_j(\bv) \psi(v_1\cdots v_j \cdots v_p \times\cdots\times) \label{eq:one_flip}
\end{equation}
for every admissible $\bv$. 

With an unnormalized state, without loss of generality, we start with the reference amplitude
\begin{equation}
    \psi(0, 0, \cdots, 0, \times, \cdots, \times)\equiv 1
\end{equation}
and successively apply Eq.~\eqref{eq:one_flip} for $j=1$ up to $p$. Let $\bv_j=(v_1, \cdots, v_j, 0, \cdots, 0, \times, \cdots, \times)$, this procedure yields
\begin{equation}
\begin{aligned}
    \psi(v_1, \cdots, v_p, \times, \cdots, \times)=&\prod_{j=1}^p \big[c_j(\bv_{j-1})\big]^{v_j}\\
    =& \exp\Big(\sum_j v_j\log\alpha_j+i\pi\sum_{k<j}C_{jk}v_jv_k\Big) \label{eq:X_full}
\end{aligned}
\end{equation}
In other words, every time a bit $v_j=1$, we pick up a phase factor $c_j$. Eq.~\eqref{eq:X_full} completely specifies the unnormalized wave function on all basis states that survive the Z-type constraints, and is already in the general form of a Boltzmann machine. 

Combining Eq.~\eqref{eq:X_full} with Eq.~\eqref{eq:Z_full}, we obtain the full expression of the (unnormalized) wave function: 

\begin{equation}
\begin{aligned}
    \psi(v_1, v_2, \cdots, v_n)=& \exp\Big(\sum_{j=1}^p v_j\log\alpha_j+i\pi\sum_{k<j}C_{jk}v_jv_k\Big)\\
    \cdot & \prod_{j=p+1}^n\Big(1 + \exp(\log \alpha_j + i\pi\sum_k E_{jk}v_k)\Big)\\
    \ketPsi=& \sum_{\bv}\psi(\bv)\ketv
\end{aligned}
\end{equation}

In this procedure, we introduced terms like $\exp(i\pi v_j)$ and $\exp(i\pi v_j v_k)$. In the RBM representation, the former simply corresponds to setting the bias for the visible neuron $v_j$, and the latter means that we introduced a connection between visible neurons $v_j$ and $v_k$. Using the conclusion in \cite{gao2017efficient}, this is corresponding to adding a hidden neuron that connects to $v_j$ and $v_k$, with the connection weights computed from Eq. (\ref{eq:vconn}):

\begin{align}\label{eq:vconn}
\exp(J v_j v_k)=\sum_h \exp(&a-\ln 2+b (v_j+v_k)(2h-1)\nonumber\\
                            &+c(2h-1)+d(v_j+v_k))
\end{align}

One solution is:
\begin{equation}\label{eq:sol}
a=-d=-J/2, b=-c=-i\arccos(e^{J/2})
\end{equation}

In this way, we have finished the construction of the logical $Z$ eigenstate of an arbitrary stabilizer group, and the eigenstate of other logical operators can be constructed in the same way. The number of hidden neurons is at most $p(p-1)/2+r$, meaning that the representation is efficient. In summary, our method can be organized into Algorithm \ref{alg:RBMstabilizer}.

\begin{Example}[ {[[5, 1, 3]] code}]
{As an example, we take the $[[5, 1, 3]]$ code, the smallest quantum error correcting code that can correct an arbitrary single qubit error \cite{laflamme1996perfect}, }to illustrate the construction procedure. The stabilizer generators are:
\begin{equation}
\begin{array}{cccccc}
T_1=&X&Z&Z&X&I\\
T_2=&I&X&Z&Z&X\\
T_3=&X&I&X&Z&Z\\
T_4=&Z&X&I&X&Z
\end{array}
\end{equation}

 \begin{figure}[t]
			\centering
			\includegraphics[width = 0.4\textwidth]{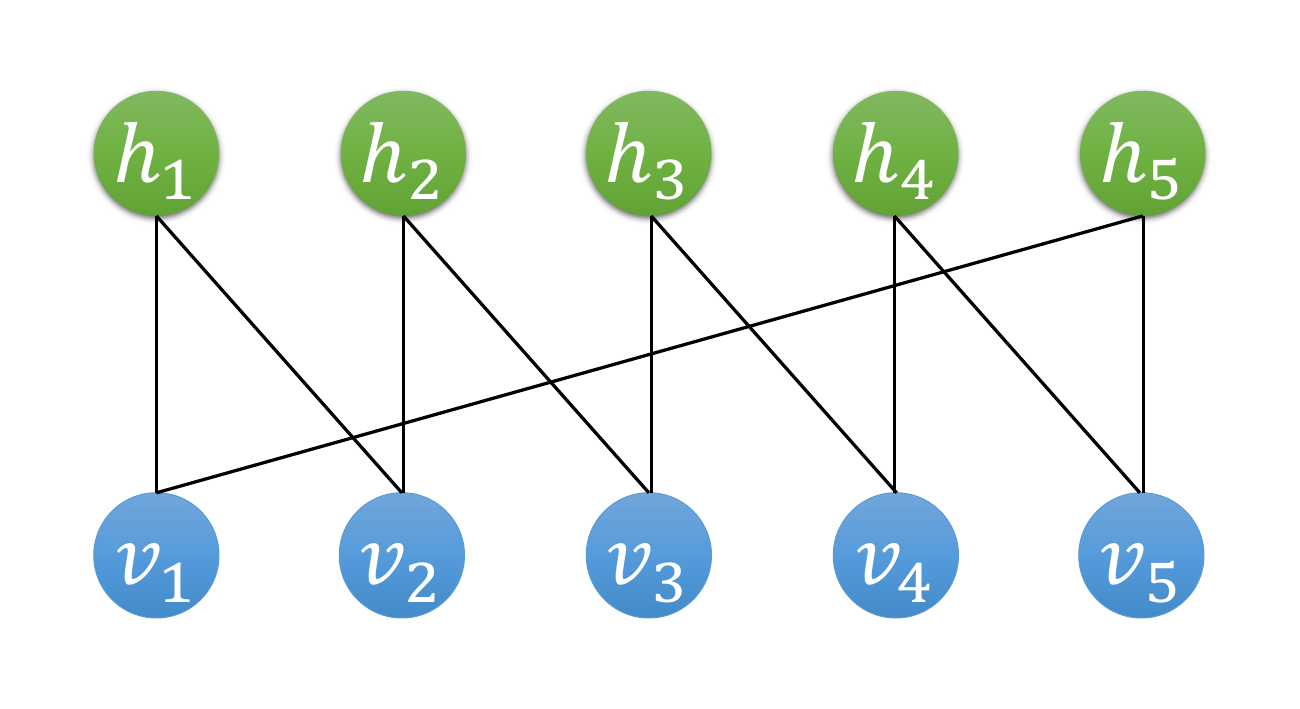}
			\caption{The RBM representation of the logical X eigenstate of the [[5,1,3]] code}
            \label{fig:513}
\end{figure}         

After Gaussian elimination, the stabilizer generators become:
\begin{equation}
\begin{array}{cccccc}
T_1=&Y&Z&I&Z&Y\\
T_2=&I&X&Z&Z&X\\
T_3=&Z&Z&X&I&X\\
T_4=&Z&I&Z&Y&Y
\end{array}
\end{equation}

Without loss of generality, we construct the eigenstate for the logical $X$ operator with eigenvalue $1$. The logical $X$ operator $\bar{X}=ZIIZX$. Since $\bar{X}|\Psi\rangle=|\Psi\rangle$, treating $\bar{X}$ as the fifth stabilizer operator $T_5$ and further carry out Gaussian elimination using $T_5$, we obtain the final form of the stabilizers:
\begin{equation}
\begin{array}{cccccc}
\tilde{T}_1=&X&Z&I&I&Z\\
\tilde{T}_2=&Z&X&Z&I&I\\
\tilde{T}_3=&I&Z&X&Z&I\\
\tilde{T}_4=&I&I&Z&X&Z\\
\tilde{T}_5=&Z&I&I&Z&X
\end{array}
\end{equation}

There are no Z-type stabilizers, and $\Psi(v_1v_2v_3v_4v_5)$ is obtained by the procedure specified in Algorithm~\ref{alg:RBMstabilizer}. Explicitly writing down every term during the transition from $|00000\rangle$ to $|v_1v_2v_3v_4v_5\rangle$, we obtain:
\begin{align*}
&\psi(v_1v_2v_3v_4v_5)\\
=&\exp(v_1\cdot 0)\exp(v_2\cdot i\pi v_1)\exp(v_3\cdot i\pi v_2)\\
\times&\exp(v_4\cdot i\pi v_3)\exp(v_5\cdot i\pi (v_1+v_4))\\
=&\exp\Big(i\pi(v_1v_2+v_2v_3+v_3v_4+v_4v_5+v_5v_1)\Big)
\end{align*}

{Explicitly converting the visible connections to hidden nodes using Eq.~\eqref{eq:sol}, the result is: }

\begin{equation}
\begin{aligned}
    &\psi(v_1v_2v_3v_4v_5)=\exp\left((i\pi+\alpha)\sum_{i=1}^{5}v_i\right)\\
    &\times\prod_{i=1}^5\Big(1+\exp\left(i\pi+2\alpha)(1-v_i-v_{i+1})\right)\Big),
\end{aligned}
\end{equation}
{where $\alpha=\log(1+\sqrt{2})$ and $v_6=v_1$. A constant factor is omitted. }

The structure of the RBM is shown in Fig. \ref{fig:513}.
\end{Example}

\begin{figure}[t]
    \centering
    \includegraphics[width=0.5\textwidth]{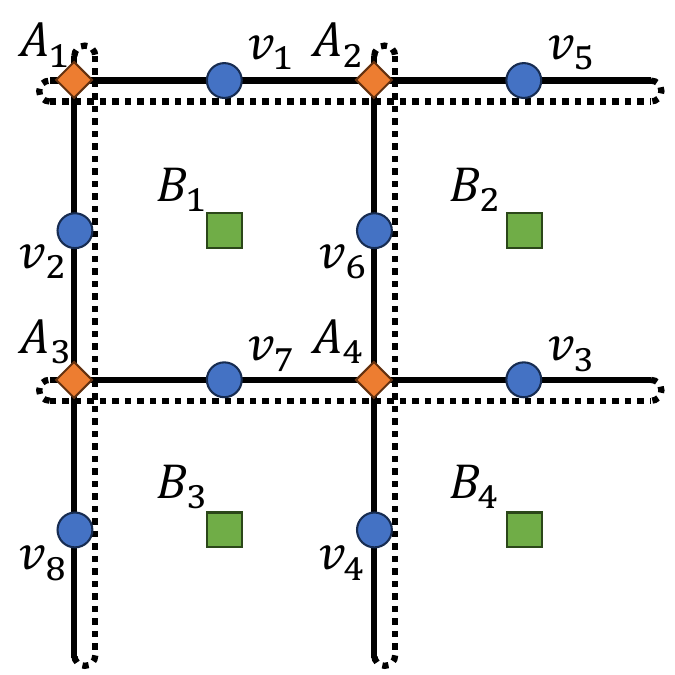}
    \caption{The toric code defined on a $2\times 2$ lattice with periodic boundary condition. The indices of the spins are defined such that no column swapping is required during Gaussian elimination. 
    \label{fig:toric}}
\end{figure}

\begin{Example}[Toric code] The toric code is a paradigmatic example of a topological quantum error-correcting code, playing a central role in quantum memory, topological quantum computation, and the study of topological phases of matter~\cite{Kitaev2003}. In this section, we demonstrate how to obtain the RBM representation of its code states using our proposed algorithm.
For simplicity, we consider the toric code defined on a $2 \times 2$ square lattice with periodic boundary conditions (i.e., a toric code on a torus); see Fig.~\ref{fig:toric}. In this setting, there are four vertex operators $A_v$ and four plaquette operators $B_p$:

\begin{equation}
\begin{array}{llll}
    A_1 = X_1 X_2 X_5 X_8,\quad &A_2 = X_1 X_4 X_5 X_6,\quad &A_3 = X_2 X_3 X_7 X_8,\quad &A_4 = X_3 X_4 X_6 X_7\\
    B_1 = Z_1 Z_2 Z_6 Z_7,\quad &B_2 = Z_2 Z_3 Z_5 Z_6,\quad &B_3 = Z_1 Z_4 Z_7 Z_8,\quad &B_4 = Z_3 Z_4 Z_5 Z_8
\end{array}
\end{equation}

One can check that $A_1 A_2 A_3 A_4=I$ and $B_1 B_2 B_3 B_4=I$, indicating that two operators are redundant. After Gaussian elimination, the independent stabilizer generators are: 

\begin{equation}
\begin{array}{lll}
    T_1 = X_1 X_4 X_5 X_6,\quad &T_2 = X_2 X_4 X_6 X_8,\quad &T_3 = X_3 X_4 X_6 X_7\\
    T_4 = Z_3 Z_4 Z_5 Z_8,\quad &T_5 = Z_2 Z_4 Z_6 Z_8,\quad &T_6 = Z_1 Z_4 Z_7 Z_8
\end{array}
\end{equation}

As an example, we construct the simultaneous eigenstate of the logical operators $\bar{X}_1=X_4 X_8$, $\bar{Z}_2=Z_3 Z_7$ with eigenvalue 1. Treating $\bar{X}_1$ and $\bar{Z}_2$ as additional stabilizer operators and further carrying out Gaussian elimination, we obtain

\begin{equation}
\begin{array}{llll}
    \tilde{T}_1 = X_1 X_5 X_6 X_8,\quad &\tilde{T}_2 = X_2 X_6,\quad &\tilde{T}_3 = X_3 X_6 X_7 X_8,\quad &\tilde{T}_4 = X_4 X_8\\
    \tilde{T}_5 = Z_1 Z_5,\quad &\tilde{T}_6 = Z_1 Z_2 Z_3 Z_6,\quad &\tilde{T}_7 = Z_3 Z_7,\quad &\tilde{T}_8 = Z_1 Z_3 Z_4 Z_8
\end{array}
\end{equation}

Following Algorithm~\ref{alg:RBMstabilizer}, $\tT_1$ through $\tT_4$ do not introduce any term into the RBM, while $\tT_5$ through $\tT_8$ each introduces a hidden neuron. Explicitly writing out all the terms, we obtain:
\begin{equation}
\begin{aligned}
    \psi(\bv)=&\Big(1+\exp\big(i\pi(v_1+v_5)\big)\Big)\cdot\Big(1+\exp\big(i\pi(v_1+v_2+v_3+v_6)\big)\Big)\\
    \cdot&\Big(1+\exp\big(i\pi(v_3+v_7)\big)\Big)\cdot\Big(1+\exp\big(i\pi(v_1+v_3+v_4+v_8)\big)\Big)
\end{aligned}
\end{equation}

The structure of the RBM is shown in Fig. \ref{fig:RBM_toric}.

 \begin{figure}[t]
			\centering
			\includegraphics[width = 0.6\textwidth]{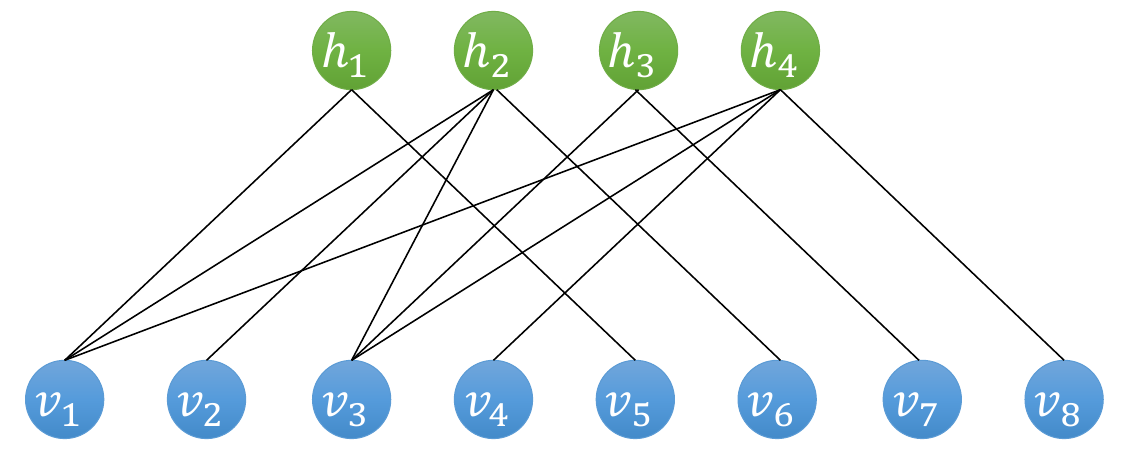}
			\caption{ The RBM representation of one eigenstate of the toric code on a $2\times 2$ torus lattice. 
            \label{fig:RBM_toric}}
\end{figure}         

\end{Example}

\section{Conclusions and Discussions}

In this work, we systematically investigate the RBM representations of stabilizer code states and present an algorithmic procedure to construct the RBM parameters for a given stabilizer group $\mathbf{S}$. To the best of our knowledge, this constitutes the first rigorous proof that RBMs can exactly and efficiently represent arbitrary stabilizer code states. 
While several related works (e.g., ~\cite{pei2021compact,Chen2025RBM}) have emerged since our initial preprint appeared on arXiv in 2018 (see also earlier work~\cite{jia2018efficient,Deng2017a,Deng2017}, where certain special cases were addressed), this topic remains under active investigation.

Our results provide new insights into the representational power of RBMs and offer a theoretical foundation for their empirical success in modeling highly entangled quantum states. Furthermore, given the central role of stabilizer codes in quantum error correction, our work opens up new possibilities for the classical simulation of quantum error-correcting codes using RBMs, offering a practical tool for initializing code states.

Despite the progress that has been made, several crucial directions remain open for further investigation:
\begin{enumerate}
    \item In this work, we presented the construction for $\mathbb{Z}_2$ stabilizer codes. However, its generalization to $\mathbb{Z}_N$ (or more generally, to finite groups~\cite{Kitaev2003,Beigi2011the,Cong2017} and to the (weak) Hopf algebra setting, see e.g.,~\cite{Buerschaper2013a,jia2023boundary,Jia2023weak,jia2024generalized,jia2024weakhopfnoninvertible,jia2024quantumcluster,chen2021ribbon,meusburger2017kitaev,jia2024weakTube,jia2022electricmagnetic,chang2014kitaev}) remains largely unexplored. 
    Such a generalization is not only of interest for applications in quantum memory and error correction, but also plays an essential role in understanding quantum phases of matter described by local commuting projector Hamiltonians, where efficient descriptions of ground and excited states are highly desirable.
    \item The relationship between RBM representations and tensor network representations has attracted significant attention in recent years (see, e.g.,~\cite{Chen2018,zheng2019RBM}). It would be interesting to relate our results to existing approaches for representing quantum codes using tensor networks. Furthermore, exploring connections between RBM representations and other neural network architectures—such as convolutional neural networks, transformers, and others—is another promising direction for future research~\cite{Carleo2019machine,jia2019quantum,lange2024architectures}.
    \item Since the code state can be represented using a RBM, establishing RBM representations of quantum operations—such as measurements and quantum channels—would allow one to embed a quantum code fully within the RBM framework. This may offer new perspectives on leveraging machine learning techniques to assist with quantum error detection and correction.
\end{enumerate}

\begin{acknowledgments}
Y.-H. Zhang thanks Xiaodi Wu, Fangjun Hu and Yuanhao Wang for helpful discussions. Z. Jia acknowledges Liang Kong for discussions during his stay in Yau Mathematical Sciences Center, Tsinghua University, and he also thanks Giuseppe  Carleo and Rui Zhai for discussions during  the first international conference on ``Machine Learning and Physics" at IAS, Tsinghua university.
This paper has been available on arXiv for several years, during which we have received valuable feedback from the community. For this version, we gratefully acknowledge the anonymous referee for suggesting a substantial revision of the references to incorporate many recent developments in the field.
\end{acknowledgments}

\bibliographystyle{apsrev4-1-title}
\bibliography{Refbib}

\end{document}